\documentclass[a4,usenatbib,fleqn]{mn2e}
\usepackage{amsmath,amssymb,amsbsy}
\usepackage[pdftex]{graphicx} %
\usepackage[backref=none]{hyperref}
\usepackage[usenames,dvipsnames]{color}
\usepackage{bm}
\usepackage{txfonts}
\usepackage{amsfonts}
\usepackage{natbib}
\DeclareSymbolFont{cmletters}{OML}{cmm}{m}{it}
\DeclareMathSymbol{v}{\mathalpha}{cmletters}{"76}

\definecolor{MyDarkBlue}{rgb}{0,0.1,0.7}
\hypersetup{pdfborder={0 0 0},colorlinks,breaklinks=true,
  urlcolor={blue},citecolor={MyDarkBlue},linkcolor={magenta}}

\title[Stimulated-emission based model of FRBs]{Stimulated-emission based model of fast radio bursts}

\author[Do{\u g}an \& Ek{\c s}i]{Mustafa Do{\u g}an$^1$ and Kaz{\i}m Yavuz Ek\c{s}i$^2$ \\  
 $^1$Istanbul Technical University,
   Faculty of Electrical-Electronics Engineering,
  Department of Control and Automation Engineering, \\ 34469,  Istanbul, Turkey, \href{mailto:mustafadogan@itu.edu.tr}{mustafadogan@itu.edu.tr}\\
 $^2$Istanbul Technical University,
  Faculty  of Science  and  Letters,  Physics Engineering  Department,
  34469,  Istanbul, Turkey, \href{mailto:eksi@itu.edu.tr}{eksi@itu.edu.tr}}

\begin{document}

\maketitle

\begin{abstract}
Fast radio bursts (FRBs) are bright, short-duration radio transients with very high brightness temperatures implying highly coherent emission. We suggest that the FRBs are caused by the self-focusing of an electron beam interacting with an ambient plasma right beyond the light cylinder radius of a neutron star. The magnetic field at the light cylinder radius is relatively high which can accommodate both young Crab-like systems and old millisecond pulsars addressing the diverse environments of FRBs. At the first stage, the intense pulsed-beam passing through the background plasma causes instabilities such that the trapped particles in local Buneman-type cavitons saturate the local field. The beam is then radially self-focused due to the circular electric field developed by the two-stream instability which leads to Weibel instability in the transverse direction. Finally, the non-linear saturation of the Weibel instability results in the self-modulational formation of solitons due to plasmoid instability. The resonant solitary waves are the breather-type solitons hosting relativistic particles with self-excited oscillations. The analytical solutions obtained for non-linear dispersion and solitons suggest that, near the current sheets, the relativistic bunches are accelerated/amplified by klystron-like structures due to self-excited oscillations by the induced local electric field. Boosted coherent radio emission propagates through a narrow cone with strong focusing due to radial electric field and magnetic pinching. The non-linear evolution of solitons and the stimulated emission are associated with the Buneman instability and the possibility of the presence of nanosecond shots in FRBs are investigated.
\end{abstract}


\section{Introduction}
\label{sec:intro}

Fast radio bursts (FRBs) are bright ($\sim 0.1-1\,{\rm Jy}$) radio transients of duration $\sim 0.1-10\,{\rm ms}$ with very high brightness temperatures implying highly coherent emission \citep{lor+07,tho+13}.
Their dispersion measures (DM) being in excess of the Galactic contribution \citep{cor02}, their isotropic distribution in the sky \citep{cha+16} as well as their recent localisation \citep{cha+17,ban+19,rav+19} suggest they are cosmological sources \citep[see][for reviews]{pet+19,cor19}. 
To date $\sim 10^2$ sources are detected at frequencies ranging
between $400\,{\rm MHz} - 8\,{\rm GHz}$ \citep[see][for a catalog of FRBs]{pet+16}\footnote{See the link \url{http://frbcat.org} associated with the paper.}.
FRBs have isotropic equivalent luminosities as
high as $\sim 10^{43}\,{\rm erg\,s^{-1}}$ and energies $\sim 10^{40}\,{\rm ergs}$ \citep{tho+13}.
The most remarkable property of FRBs is the coherency of their emission as implied by their very high brightness temperatures $T_{\rm B} \gtrsim 10^{35}\,{\rm K}$.

Several exotic models  such as cosmic strings \citep{yu+14,cao18}, collisions between neutron stars and asteroids/comets \citep{gen15,dai+16,sma+19}, mergers of charged black holes \citep{zha16}, axion quark nugget dark matter model \citep{wae19}, black-to-white hole transition by non-perturbative quantum gravity effects \citep{bar+18} and collapse of magnetospheres of Kerr-Newman black holes \citep{liu+16} have been proposed as the origin of FRBs \citep[see][for a catalog of theories]{pla+18}.
The detection of repeating FRB sources \citep{spi+16,sch+16,ami+19,and+19,kum+19,mar+20,fon+20}, the deficiency of likely cataclysmic progenitors such as neutron star mergers \citep{fal14,mos+18} in supplying the occurrence rate of the FRBs per unit volume \citep{rav+19} and population studies \citep{bha+19} suggest FRBs are non-catastrophic events. The detection of faint pulses from FRB 171019 \citep{kum+19} implies that most FRBs repeat, even though they are undetected due to poor localisation.

Frequency drifts in FRBs point to neutron stars magnetospheres as the likely site for the origin of FRBs \citep{lyut19a,lyut19b}.
Two of the likely models in this category are the magnetar model \citep{lyu02,pop10,kea+12,lyub14,pen15,kat16,wan17,wan+18,wad19,cao17,bel17,bel19,maa+19} which suggests that 
FRBs are flares from young magnetars
and the super-giant pulse model \citep{kul+14,cor16,con+16,pop16,lyut+16,mun+19} which suggests that FRBs are extreme samples of giant pulses from younger-than-Crab rotationally powered pulsars.

The giant pulse model has the following advantages over the magnetar model: 
\begin{enumerate}
\item No radio emission was detected from SGR 1806-20  when it showed the magnetar giant flare \citep{ten+16}; 

\item  No FRBs were detected from the six gamma-ray burst remnants with possible magnetar engines \citep{men+19}. 
The lack of high energy emission from FRBs is consistent with the lack of any enhancement in the high energy emission of Crab pulsar during giant pulses \citep{lun+95,ali+12,bil+12,mic+12,hit+18,ahn+19}.
Detection of high energy emission contemporaneous with an FRB would strongly favour the magnetar model.

\item FRB 180814 \citep{ami+19} appears to show a $13\,{\rm ms}$ period \citet{mun+19} within its sub-pulses which implies that FRB 180814 host a rotationally powered pulsar.

\item The pulse-energy distribution of the repeating FRB121102 \citep{spi+14,spi+16}, is
well described by a power law with index $\alpha = -2.8 \pm 0.3$ \citep{gou+19}, well in agreement with what \citet{ber19}
find for the Crab giant pulses \citep[see also][]{arg72,maj+11,mic+12}. Actually, this is not a very strong argument as the power-law nature of the amplitude distribution of bursts can be explained also by the magnetar model \citep{wan17}.

\item FRBs are localized to both star forming (high-metallicity) galaxies and low-metallicity ones suggesting that they can have both young and old central engines. As magnetars are preferentially young objects they can not address the localisation of some FRBs to low-metallicity environments. Giant pulses, on the other hand, are observed both from Crab-like young pulsars and recycled millisecond pulsars and are thus capable of addressing occurrence of FRBs in both environments.
\end{enumerate}

There are problems also with the SGP model of FRBs: 
The instantaneous radio efficiency of SGPs as those seen in the Crab pulsar \citep{cor+04} can reach values as high as $\sim 10^{-2}$ \citep{cor16,lyut+16}. 
Although this is well above the typical radio efficiency of pulsars, $\sim 10^{-5}$, it is still not enough to address the cosmological distribution as implied by both the dispersion measures and the redshifts, given the SGP models are limited by the spin-down power of young neutron stars. 
Indeed the SGP model initially was proposed as an extra-galactic but non-cosmological explanation for FRBs \citep{cor16,lyut16}. Soon
after the localisation of the repeating FRBs at $\sim 1\,{\rm Gpc}$, \citet{lyut17} argued that the measured cosmological distances exclude SGP model as the origin of FRBs \citep[see also][]{mey+17}. Yet another well known issue is why more of the FRBs are not discovered from nearby galaxies but cosmological distances.

Note that FRBs must be emitted in low-density plasma as
$\nu \sim 1\,{\rm GHz}$ radiation cannot propagate through plasma with
$n_{\rm e} \sim 10^{10}\,{\rm cm^{-3}}$. This introduces another limitation for the SGP model of FRBs since the central engines of FRBs,  i.e.\ the rotationally powered pulsars, are then required to be older than 10 years, so that the supernova remnant (SNR) is transparent to GHz radio emission \citep{mey+17,bie19}.
 
Recently, \citet{mac+19self-trap} proposed a non-linear optical phenomenon, so called ``self-trapping'' \citep{chi+64} as an intrinsic ingredient to any possible model to address the rare occurrence of FRBs and the luminosities observed. 
In this work, we suggest that FRBs are emitted from the current sheets right beyond the light cylinder radius of neutron stars (see \S \ref{sec:model}) by the self-focusing electron beams composed of relativistic bunches with self-excited oscillations. For the first time to our knowledge we suggest that these oscillations are induced by breather-type solitons  within the context of the stimulated emission providing the non-linear coherent radiation. The narrow beaming of the emission allows them to be observed from much larger distances with the inferred isotropic equivalent luminosities. We propose a model of two concentric cylinders where the outer one is the relatively slow ambient plasma and the inner one is the relativistic plasma jet\footnote{This is inspired by terrestrial maser production at room temperature \citep{breeze+18}.}. The observed narrower pulse width of the emission corresponds to smaller beam opening angles \citep{ZhangF18}. We find that FRBs consist of nanosecond shots such as seen in SGPs \citep{cor+04}; their angle of aperture thus can be as small as $\theta \approx 3 \times 10^{-8}$ radians or the opening angle for the coherent emission can set the width of the beam. Our antenna-like maser emission model, proposed in the next section, is different from the  maser emission models in the literature \citep{cordes+19,ZhangF18,Asseo+80,plo19} in that we explain the population inversion dynamics.
Finally, in \S~\ref{sec:discuss} we discuss the implications of our model for FRBs.

\section{A mechanism for FRBs}
\label{sec:model}

It is well established that the electric fields generated by a rotating neutron star is strong enough to extract charged particles from the surface of the star \citep{gol69}. This forms a corotating plasma in the magnetosphere with density of the particles $n_{\rm GJ} \simeq 7\times 10^{10} B_{12}/P\, {\rm cm^{-3}}$ where $B_{12}$ is the surface magnetic field in units of $10^{12}\,{\rm G}$ and $P=2{\rm \pi}/\Omega$ is the rotation period of the pulsar \citep{gol69}.  The corotation of the particles is not possible beyond the light cylinder radius $R_{\rm L} \simeq 5\times 10^9 P\, {\rm cm}$ and accordingly the field lines crossing $R_{\rm L}$ open. The open field lines converge on the surface of the star to the polar caps with an opening angle $\theta_{\rm p} \simeq \sqrt{R/R_{\rm L}} \simeq 0.014 P^{-1/2}\, {\rm rad}$. The secondary electron-positron pairs created by the $\gamma$-rays \citep{rud75} are accelerated along the open field lines emitting curvature radiation. 

The dipole magnetic field of the neutron star declines with the radial distance $r$ as $B = \mu/r^3$ where $\mu$ is the magnetic dipole moment. The magnetic field at the light cylinder radius, $B_{\rm L} = \mu/R_{\rm L}^3$ is then
\begin{equation}
B_{\rm L} = (2\pi)^3 \mu /c^3 P^3
\label{BL}
\end{equation}
Assuming the neutron star is spinning down under magnetic dipole torques in a corotating plasma \citep{spi06}
\begin{equation}
I \frac{{\rm d} \Omega}{{\rm d}t} = - \frac{\mu^2 \Omega^3}{c^3} (1 + \sin^2 \alpha)
\end{equation}
where $I$ is the moment of inertia and $\alpha$ is the inclination angle between rotation and magnetic axis. Solving the magnetic dipole moment from this equation we obtain
\begin{equation}
\mu = \frac{1}{2\pi} \sqrt{\frac{c^3I}{1+\sin^2 \alpha}} \sqrt{P\dot{P}}
\label{mu}
\end{equation}
Using \autoref{mu} in \autoref{BL} we find
\begin{equation}
B_{\rm L} = (2\pi)^2 \sqrt{\frac{I\dot{P}}{c^3P^5(1+\sin^2 \alpha)}}  \simeq 2 \times 10^8 \,{\rm G}\, \dot{P}^{1/2} P^{-5/2}
\label{B_lc}
\end{equation}
where we assumed $I = 10^{45}\,{\rm g\, cm^2}$ and $\alpha=45^{\circ}$.

The magnetic field at the light cylinder radius can be high ($B_{\rm L} \sim 2-3\times 10^5$~G) not only for Crab-like young neutron stars ($P=0.033$~s, $\dot{P} = 4.2\times 10^{-13}$~s/s) but also for millisecond pulsars ($P\sim 2 \times 10^{-3}$~s, $\dot{P} \sim 10^{-20}$~s/s) which are old systems. Giant pulses tend to originate from pulsars with strong $B_{\rm L}$ \citep[e.g.][]{wan+19gp} and are observed in millisecond pulsars as well as young Crab-like systems.  The magnetar model of FRBs to date, however, can only address young systems\footnote{Of course a short lived magnetar can form in old systems by merger of two neutron stars. These are associated with short gamma-ray bursts. Such a progenitor can not be associated with FRBs given that their rate of occurrence is much lower than FRBs and obviously that such a onetime process can not address repeating FRBs.}. We emphasise this point as FRBs are observed in diverse environments of both low-metallicity and high-metallicity \citep{mar+20} indicating that they originate in both old and young systems. We note also that the period of a millisecond pulsar does not change as rapidly as a younger-than-Crab pulsar which implies systems that can live longer.

\subsection{Early stages of the instability}

\begin{figure*}
\begin{center}
    \includegraphics[width=2\columnwidth]{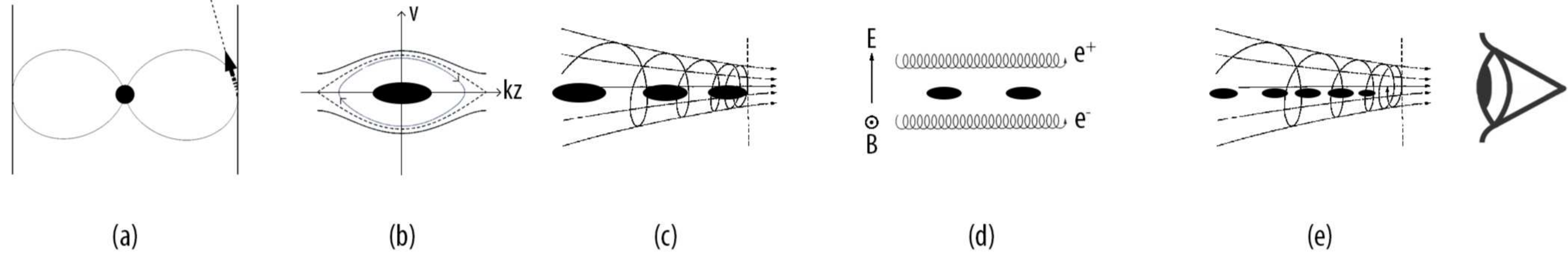}
  \end{center}
  \caption{Illustration of the steps of the proposed mechanism:
(a) the neutron star (the solid circle at the center) and its magnetosphere bounded by its light cylinder (the vertical solid line), seeding the coherent radiation (in the direction of arrow);
(b) the creation of the bunched beam, within the trapping region, by the Buneman instability in phase space;
(c) the mirroring, acceleration and focusing region by the two-stream instability with radial electric and non-uniform magnetic fields;
(d) the bunched beam gains more kinetic energy by $\mathbf{ E\times B}$ drift; 
(e) the focusing and boosting of the coherent emission as maser by velocity modulation along the line of sight.}
\label{fig:main}
\end{figure*}

The model we propose here is motivated by the plasma-beam systems in terrestrial experiments (see \autoref{fig:main}).
Plasma-beam systems have many applications in fusion reactors,
astrophysics and plasma wake-fields for accelerating electrons to
higher energies.
These systems initially are both charge and current naturalised. 
The initial charge neutralisation implies the quasi-neutral property of plasma, while current neutralisation requires at least two streaming channels. 
Streams are often considered to be in opposite directions, but they can be parallel with opposite signs of currents. 
An intense pulsed-beam propagating through an ambient plasma leads to the development of streaming instabilities \citep{chen16}.  
In nearly collisionless (low density) plasma, such instabilities based on kinetic behaviour of plasma, e.g.\ Buneman, two-stream, LHDI (lower hybrid drift instability) and Weibel, are known to cause particle trapping and acceleration in a relativistic setup \citep{kraso08,Bret+09,chen16,tok+18}. At the initial stage, in plasma excited with a beam of bunched particles, the Buneman instability develops \citep{che+09}. 
\citet{gal+81} proposed that, at the non-linear stage of the the Buneman instability, short-lived non-stationary double layers or charge-separated regions will form by pinched electron currents due to strong longitudinal magnetic fields. At the boundaries of these regions, local electrons are accelerated excessively by the high electrical potential. The charge-separated regions thus become densely populated \citep{gal+81}. The trapping of the electrons by the Buneman instability leads to the growth of new instabilities, such as two-stream instability, following the non-linear saturation \citep{che+09}. We require that the plasma density does not change smoothly at short distances so that the particles are decelerated and trapped in cavitons by the Buneman instability. 

This saturation---or other possible ones due to sustained short duration oscillations, e.g.\ bouncing in small trap or caviton---can be disrupted by self-excited or triggered plasma events\footnote{Such  that the growth  of the Buneman instability is assisted by the self-excitation for sufficiently small time cycles \citep{gal+81} and likewise plasma triggering is observed in microwave pulse compressors built to obtain narrow high peak pulses \citep{kars+19}. The following plasma behaviour such as new matched oscillations can result in higher amplitude pulse formation.}. For example, the two-stream kinetic instability occurs when there are counter-streaming plasma flows in the velocity space in the presence of radiation pressure. The two-stream instability driven plasma can create a radial electric field, similar to the ponderomotive force dynamics (caviton formation, beam focusing, wave compression) \citep{tok+18}. If the two-stream instability is excited, then the self-focusing of an electron beam moving through a plasma can be observed. This radial self-focusing is similar to a beam passing through a channel in a plasma. The radius of the beam, hence, diminishes in time while the local density of the particles increases uniformly \citep{kraso69}. The plasma, thus, acts as a non-linear medium for the focusing of the trapped particles \citep{mac+19self-trap}. The efficiency of focusing increases at the relativistic regime (see \S~\ref{sec:discuss}). 

Following the two-stream instability, Weibel instability will develop in the transverse direction \citep{nishi+07}. 
Weibel-like (filamentary) instabilities are known to be the underlying cause of the transverse field growth in a plasma \citep{chen16}.
The Weibel instability and the main electron acceleration are stimulated in the downstream region of electron-positron jets with plasma density perturbations leading to the formation of current filaments \citep{nishi+07}. 
In the relativistic regime, the kinetic energy  of these filaments is transferred to the magnetic field. 
The energy stored in the magnetic field is transferred partially back to the plasma particles due to saturation. 
The non-linear saturation developed at this stage is the origin of the induced electrostatic field which is responsible for the redistribution of particles along with the help of non-linearities, e.g.\ relativistic mass variation and ponderomotive force \citep{ghiz+13,kars+19,farin+04}. 

\subsection{Non-linear evolution of the solitons and population inversion}
\label{ssec:popinv}

The Weibel instability is an electromagnetic instability, that arises from plasma anisotropy. The Weibel instability generates a magnetic field which is perpendicular to the direction of the anisotropy which is reduced by axial momentum transfer. In a thin current sheet, seed-X-points are generated by the Weibel instability. The inner current region decays into a ``magnetic vortex street'' consisting of plasmoids and seed-X points \citep{treu+10,com+17}.  The importance of the plasmoid instability in actuating fast reconnection and fast energy transfer has recently gained attention \citep[see][for a review]{kag+15}. The role of plasmoids in the pulsar emission mechanism has recently been established \citep{cer17,phi+19,lyub19}. The plasmoid instability can be explained in terms of a tearing instability occurring in a reconnecting current sheet. This tearing process will force the inner current sheet to decay into a chain of highly dynamical magnetic islands with meso-scale plasmoids \citep{treu+10,com+17}. These plasmoids can be depicted by the thin magnetic braids in the central part of the current sheet \citep{treu+10,com+17,zen+08}.

The plasma, when a powerful electromagnetic wave propagates within, acts as a non-linear medium. The high amplitude wave leads to the anisotropy of the medium resulting in further enhancement of the electric field and hence to the growth of the refraction index which is the origin of the non-linear dispersion \citep{mac+19self-trap}. As the plasma wave is non-linear the ponderomotive force of the plasma waves removes the background plasma. Then a local depression in density constitutes a caviton. Plasma waves trapped in this cavity then form an isolated structure called an envelope soliton \citep{chen16}.

During the rearrangement of the magnetic field topology, plasmoid instability evolves into current sheets intermittently that have Langmuir plasmons. These high-frequency plasmons can form Langmuir solitons due to modulational instability \citep{liandzhang97, chen16}. Solitons or solitary waves propagating in non-linear dispersive media ``pass-through'' one another without losing their identity \citep{ZabandKru65}. 
Non-linear dispersion can cause compression/focusing or generation of solitons, unlike the linear case.
The electrostatic fluctuations and scattering cause current disruption in the central region of the current sheet. The non-uniform current disruptions re-create the magnetic reconnection, namely merging the magnetic field lines \citep{arons11,singh04}. Thus, the magnetic coalescence at X-points results in the conversion of magnetic energy to kinetic energy and particle acceleration. The fastest acceleration during magnetic reconnection occurs at the initial catastrophic X-point collapse, with the reconnection of electric fields. 
During the X-point collapse, particles are accelerated by charge-starved excessive electric fields resulting in immediate arc-like discharges and pinched currents due to plasmoid instability \citep{lyutikov+16}. The nature of the interaction between beam particles (bunched together) and electromagnetic field while the interaction (energy-exchange) is significantly enhanced by the relativistic regime, explains the conditions for the existence of resonant solitons in non-equilibrium plasma beam systems. The energy density of the resonant electromagnetic solitons in non-equilibrium dispersive media is preserved  \citep{Bachin+80}. 

To understand the coherent microwave emission, antenna-like maser mechanism with resonant peaks \citep{cordes+19,ZhangF18,Asseo+80,plo19}  with non-monotonic charge distribution due to electromagnetic trapping in plasma should be considered. Assuming plasma creation is dominated by the pair production caused by high energy photons, electron and positron densities can be similar. In our model, the self-excited maser mechanism can be explained such that the inner cylinder, as the magnetic mirror/bottle bounded by circular electric fields, is surrounded by ambient plasma which has non-uniform magnetic field.  Here the inner cylinder represents the sapphire cavity and the outer one represents the copper cavity in terrestrial maser setup \citep{breeze+18}. 
To grasp the coherence of the emission, klystron working principles can be summarised as follows: After the bunching of the electrons by the induced electric field, the bunches drifting by $\mathbf{E \times B}$, rather than the thermal one,  gain more energy for a while; the high kinetic energy of the bunches is then transferred to the electromagnetic wave by magnetic coupling or reconnection (see \autoref{fig:main}). 
This charge separation due to radiation pressure (longitudinal electric fields with the strong external magnetic field) causes the non-linear saturation of the field amplitude similar to kinetic instabilities, e.g.\ the inherited Buneman, two-stream and Weibel instabilities in the current sheets \citep{arons11,singh04}. 

The large amplitude electromagnetic pulses propagating in a plasma along a strong magnetic field under cyclotron resonance conditions are shown to take the form of solitons \citep[][\S 5.5]{kraso08}. The auto-resonance of solitons is then developed by the acceleration of the electrons to ultra-relativistic speeds suppressing the cyclotron resonance.
Consequently, the coherent emission should be concentrated in a narrow hollow cone in the vicinity of the current sheets' boundary. The cone must be hollow since the radiation is generated only near the current sheets, thus tracking the last open field lines (LOFL). The cone must be narrow because the opening angle is set by the nearly vertical directions of the LOFL above the polar caps \citep{ZhangF18}. The observed radio signals from pulsars are quite narrow in time. For example, the first pulsar discovered, PSR B1919+21, has a period of 1.34 s but a pulse width of only around 40~ms which translates into a beam opening angle of one-tenth of the pulsar inclination angle \citep{ZhangF18}.

The non-linear envelope solitons as self-modulational instability (caused by relativistic mass variation and ponderomotive dynamics) of pulsar micro-structures as proposed by \citet{chian+83} are stable against longitudinal perturbations and mutual collisions. The intermittent and quasi-periodic nature of the observed pulsar micro-structures/pulses can be explained by the collection of envelope solitons with randomly fluctuating amplitudes (intermittency) or the latter (quasi-periodicity), a sequence of envelope solitons with little variation in their peak amplitudes \citep{chian+83}. 
Since a two-level system is usually unstable in the presence of small electromagnetic perturbations, a coherent emission of an electromagnetic wave is indispensable due to population inversion phenomena among the levels, such as maser mechanism. Because, the ambient plasma and current sheets are free from crystal structures similar to the diamond core in terrestrial maser setup \citep{breeze+18}, we need to explain the population inversion mechanism in the current sheets. Buneman instability in current sheets results in bunching, magnetic reconnection, bifurcation of streams \citep{singh04}. The relativistic bunches are decelerated and trapped in cavitons as small resonators. Along with this process an electron velocity gap (the trapping width) of magnitude
\begin{equation}  
v_{\rm tr} = \sqrt { \frac{2 e E}{m_{\rm e} k}} \propto \frac{\omega_{\rm b}}{k} \, ,
\label{eq:trapwidth}
\end{equation}
is developed \citep{Dieckmann+00,liu87}.
Here $-e$  is the electron charge, $E$ is the amplitude of electric field at which the Buneman wave traps a significant fraction of the electron population, $m_{\rm e}$ is electron rest mass, $k$ is the wave-number and $\omega_{\rm b}$ is the bouncing frequency. During the evolution of the Buneman instability, the motion of the particles, governed by harmonic oscillator dynamics with bouncing frequency $\omega_{\rm b}$, is trapped in a travelling wave. On the other hand, the bounce-time $\propto \omega_{\rm b}^{-1}$ for the trapped-particles is increased by the lower bouncing frequency associated with the diminished amplitude in the phase space for trapping (see \autoref{fig:main}). For lower velocity electrons which have almost zero velocity at the boundary (separatrix) of the trapped region, the trapped fraction ($n_{\rm tr}$, trapped density divided by the electron density, $n_{\rm e}$)  of the population is given by \citep{Dieckmann+00,liu87}
\begin{equation}  
\frac{n_{\rm tr}}{n_{\rm e}} = \frac{1}{\sqrt{2\,{\rm \pi}}} \int_{v_{\rm p} -v_{\rm tr}}^{v_{\rm p} + v_{\rm tr}}   {\rm e}^{-v^2/v_{\rm e}^2} \, {\rm d}v  \, .
\label{eq:trap}
\end{equation}
Here  $v_{\rm p}$ is phase velocity of the wave, and $v_{\rm e} \cong  v_{\rm p} - v_{\rm tr}$ where large numbers of electrons are trapped \citep{Dieckmann+00}. Note that, the trapped fraction is exponentially sensitive to the magnitude of the electric field which is rapidly evolving in time. The population of the trapped/low-velocity electrons which can be represented as ground state carriers is increased by the population inversion followed by the stimulated emission. After being excited by the dense-beam, this population inversion is destroyed by the high-velocity bunches accelerated at the boundary as a precursor of the two-stream instability. Thus, the field-dependent trapped fraction of the population evolves in time to provide coherent stimulated emission. By this instability, broadened/enhanced electron velocity distribution (plateau-like distribution or double-humped electron distribution \citep{chen16}) establishes the oscillations in the phase space. Finally, the population inversion is achieved repeatedly around the quasi-equilibrium point of the bunch momentum \citep{zhele+01}. These cavitons are re-filled up continuously following the energy interchange between the particles and the coherent electromagnetic waves. This phenomenon of the continuous passage of plasma through the cavitons guarantees that the population inversion is kept steady \citep{plo19,chen16}. 

\subsection{Non-linear two-level system}
\label{ssec:nonlintwo}

\begin{figure*}
\centering
\includegraphics[width=0.98\columnwidth]{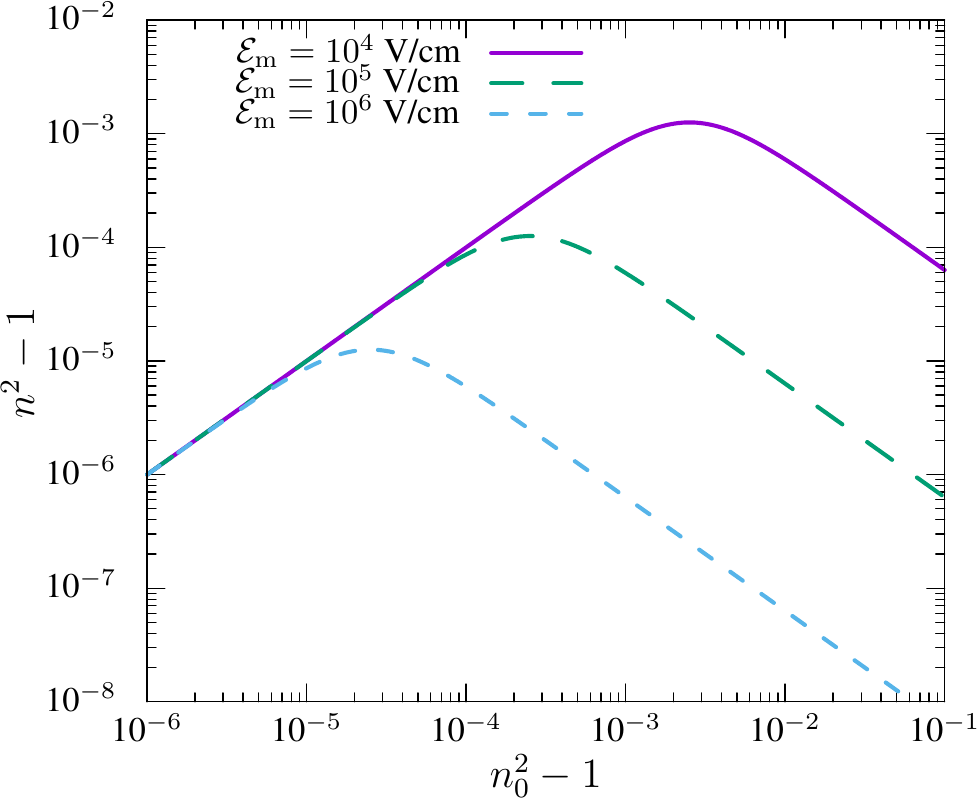}
\includegraphics[width=0.98\columnwidth]{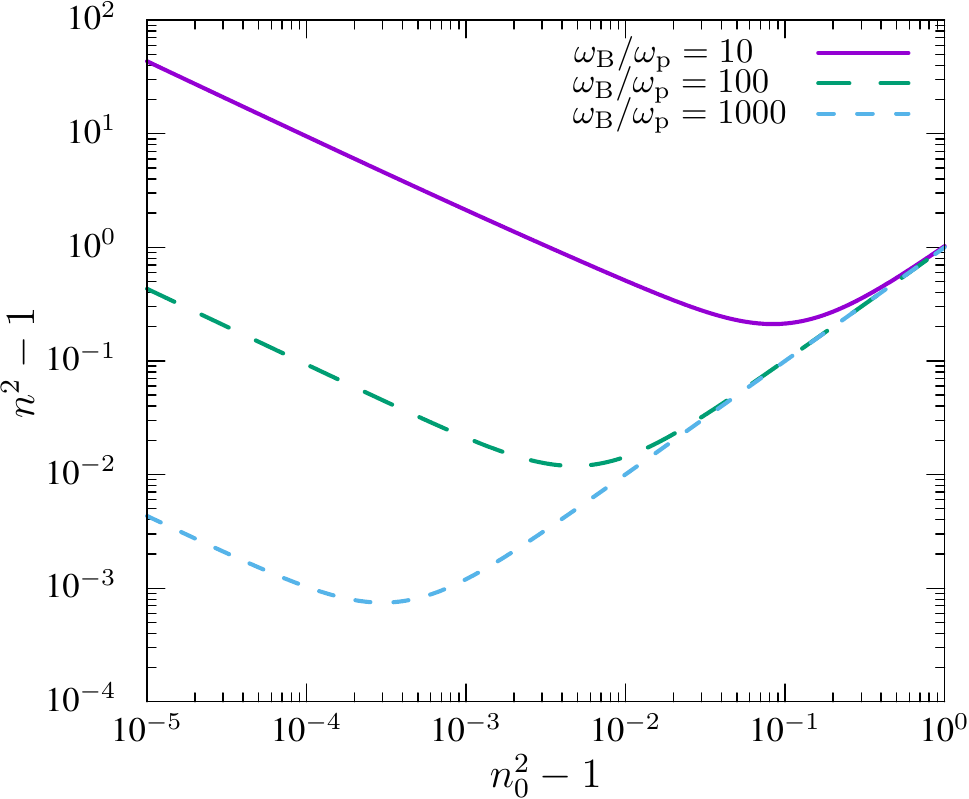}
\caption{The electromagnetic non-linearity ($n^2-1$) as a function of $n_0^2 -1$. The left panel is for the case without cyclotron resonance (auto) given in \autoref{eq:n1} and the right panel is for the case  with cyclotron resonance given in \autoref{eq:n2}.}
\label{fig:nonlin}
\end{figure*}

The resonant solitary waves in a non-linear two-level system \citep{kraso08} have  governing equations and solutions similar 
those given in \citet{chian+83} \citep[see also][]{Herve19,Sazo18,Pak16}. 
Resonant solitons are the breather-type solitons such as  those observed in non-linear optical medium \citep{kraso08,Herve19}. The non-linear dispersion law and the shape of the envelope solitons are governed by the following non-linear equations \citep[see][\S 5.5]{kraso08}:
\begin{align}
&\frac{\partial^{2} \mathbf{E} }{\partial z^2} - \frac{1}{c^2}\frac{\partial^{2} \mathbf{E} }{\partial t^2} = \frac{4 {\rm \pi}}{c^2}\frac{\partial^{2} \mathbf{P} }{\partial t^2} \label{eq:main1}\\
&\frac{\partial^{2} \mathbf{E} }{\partial t^2} + \omega_{\rm p}^2 \, \mathbf{P} = -\frac{2N d_0^2 \omega_{\rm p}}{\hbar} W \,  \mathbf{E} \label{eq:main2}
\end{align}
where $\mathbf{E}$ and $\mathbf{P}$ are the electric field and polarisation vectors, $c$ is the speed of light, $\omega_{\rm p}$ is the natural frequency of the medium (ambient plasma), $N$ is the particle density and $d_0$ is the dipole moment of separated charges within the Debye volume filled with aligned dipoles \citep{mac+19self-trap}, and the population difference for the two levels is given by \citep{kra65}
\begin{eqnarray}\label{eq:popdiff}
W = - \sqrt{1-\left(\frac{P}{N d_0}\right)^2-\left(\frac{1}{N d_0\omega_{\rm p}}\frac{\partial P}{\partial t}\right)^2}\, .
\end{eqnarray}
(see \autoref{sec:append}). In the linear approximation to the governing equations, the electromagnetic waves that have harmonics at $\exp({\rm i}\omega t-{\rm i}kz)$ (${\rm i} \equiv \sqrt{-1}$) have the following dispersion equation:
\begin{equation}
n_0^2 (\omega) \equiv \left(\frac{ck}{\omega}\right)^2 = 1 +\frac{q^2}{\delta}, \qquad \mbox{linear limit}
\end{equation}
where $\delta$ is the relativity parameter and $q$ is the energy quantisation parameter given as
\begin{equation}
\delta \equiv \frac{\omega_{\rm p}^2} {\omega^2} -1, \qquad  q^2 \equiv \frac{8 {\rm \pi} N d_0^2}{\hbar\omega_{\rm p}}.
\end{equation}
Here $\hbar\omega_{\rm p}$ is unit value of energy quantisation for the incident photons.
The parameters above should satisfy the conditions
\begin{equation}
q^2 \ll 1, \qquad |\delta| \ll 1, \qquad |n^2-1| \ll 1 \, .
\label{eq:ineq}
\end{equation}
for the existence of the soliton (see \autoref{sec:app-B}).
Note that as the number of cycles per oscillatory peak amplitude increases, attained at the limit of $q^2 / \delta^2\rightarrow 1$, the solitary wave complies better with the quasi-periodic structures observed in micro-pulses of giant pulses \citep{cor79}. On the other hand, $\delta \rightarrow 0$ should be satisfied for the matched resonance and steady population inversion as well \citep{kraso08}. Thus, the above existence conditions for the solitary wave are met properly. Here, the strong non-linear coupling results in velocity modulation. Accordingly, the non-linear dispersion law obtained analytically without any approximation is
\begin{equation}
	n^2 (\omega) = 1 +\frac{q^2}{\delta}\left[ 1+ \left( \frac{q^2 {\cal E}_{\rm m}}{8{\rm \pi} \delta}\right)^2 \right]^{-1}, \quad \mbox{nonlinear, implicit}
	\label{eq:n0}
\end{equation}
\citep[see \S 5.6][or \autoref{sec:app-B} below]{kraso08} where $n(\omega)$ depends on ${\cal E}_{\rm m}$ which is the maximum field amplitude of the soliton, ${\cal E}$. The electromagnetic non-linearity is proportional to $|n^2-1|$. In a more explicit way, we can rewrite the above expression for the dispersion diagram (see \autoref{fig:nonlin}) for magnetized plasma with suppressed cyclotron resonance as:
\begin{equation}
n^2 - 1 = \frac{n_0^2 - 1}{ 1 + (n_0^2-1)^2 ({\cal E}_{\rm m}/8{\rm \pi})^2}, \qquad \mbox{nonlin w/o cyc}
\label{eq:n1}
\end{equation}
where $n_0 = c k/\omega$. Here, the compression/focusing effect of the nonlinear dispersion, $\propto \omega^{-6}$ implies a very narrow frequency bandwidth. 
For a resonant soliton with appropriate existence conditions, the analytical solutions for the group and phase velocities, and the breather-type soliton are as follows:
\begin{align}
	v_{\rm g} &= c \,\left(1+\frac{8N\hbar\omega_{\rm p}}{E_{\rm m}^2} \right)^{-1}, \qquad v_{\rm p}=c    \\
	{\cal E} &=  {\cal E}_{\rm m}  \mathrm{sech} \left[ \frac{d_0 E_{\rm m}}{2\hbar} \left( t- \frac{z}{v_{\rm g}}\right)\right] \label{eq:group}
\end{align}
\citep[see \S 5.6 of][or \autoref{sec:app-B} below]{kraso08} where $E_{\rm m} = N d_0 {\cal E}_{\rm m}$ is the amplitude of the non-linear electric field $E=N d_0 \mathrm{Re} ({\cal E}) \exp({\rm i}\omega t - {\rm i}kz) $, for the envelope soliton. We emphasise that for solving the non-linear governing equations given in \autoref{eq:main1} and \autoref{eq:main2} we did not make any approximations or linearisations. 

Assuming the presence of highly nonlinear cyclotron resonance and almost saturated electric field such that $|\omega -\omega_{\rm B}| \ll  \omega_{\rm B} $, we can obtain the refraction index obeying the cyclotron resonance conditions \citep[][\S 5.4]{kraso08} for the nonlinear system equations as follows
 \begin{equation}
 n^2 (\omega) = n_0^2 + \frac{2\omega_{\rm p}^2}{(\omega_{\rm B}-\omega)^2 \left(n_0^2 -1\right)^{2/3}}, \quad \mbox{nonlin w/ cyc}
 \label{eq:n2}
 \end{equation}
where $\omega_{\rm B}$ is cyclotron frequency. Here, $\omega_{\rm B}$ can be set within the interval of $2-200$ GHz \citep{lyut07} depending on relativistic mass, e.g.\ typical range of the Lorentz factor, $\gamma=5-500$ for mid-energetic electrons in pulsar magnetosphere \citep{cer+15}. Finally, we observe that ignoring the nonlinear terms in the refraction index formulas, \autoref{eq:n1} and \autoref{eq:n2}, will simplify to the known linear form \citep[see e.g.][]{lyut07} as shown in \autoref{fig:nonlin}.
The obtained solutions thus fully reflect the non-linear nature of the problem and is valid for the magnetized plasma in the relativistic regime right beyond the light cylinder radius.

\section{Discussion}
\label{sec:discuss}
In this work, we proposed a novel mechanism based on non-linear two-level system \citep{kraso08} to address FRB phenomena. 
The model relies on the non-linear plasma processes near the light cylinder of a rotationally powered neutron star.
The model can address the measured cosmological distances by showing that the emission would be significantly beamed. 
Our principal result is that an astrophysical maser production mechanism feeds the klystron-like amplification of coherent emission. 
This is characterised by ``self-excitation'', ``self-focusing'' and the highly coupled non-linear nature of the plasma medium. 
To produce a stable maser, the stimulated emission should be dominant over the excitation/absorption phenomena. 
For the relativistic bunches, the bounce-time for the trapped-particles, $\propto \omega_{\rm b}^{-1}$, 
increases with the relativistic mass (see equation \autoref{eq:trapwidth} and  \S~\ref{ssec:popinv}). 
Thus, a higher rate of beam-to-wave energy transfer is achieved with increased efficiency. 
Hence, the prevailing stimulated emission is succeeded. 

The Buneman instability produces strong electron acceleration in highly magnetised plasma where a broad range of velocity distribution is provided. 
The quench time (including the growth and the saturation regimes) of the Buneman instability can be approximated as $t_{\rm q} \approx 40{\rm \pi}/\omega_{\rm p}$ 
where $\omega_{\rm p}$ is the electron plasma frequency \citep{Dieckmann+00}. 
The length of the caviton induced by the relativistic  beam can be estimated as approximately $15 \lambda_{\rm d}$ where $\lambda_{\rm d}$ is the 
electron Debye length of the plasma \citep{viey+17}. The parameters of ambient plasma around the current sheet (partially based on the parameters 
given in \citet{ZhangF18}) are summarised in \autoref{tb:param}. 

The variation of the caviton length with respect to change in the particle velocity from relativistic speed to $v_{\rm e}$ in \autoref{eq:trap} is in the range of $0.093$ cm to $31$ cm comparable to the numeric results in \citet{Dieckmann+00}. This spatial range corresponds to time interval of $3\, {\rm ps}$ to $1\,{\rm ns}$ as the caviton passage time ($\approx t_{\rm q}$). In the neighbourhood of the current sheets, the plasma frequency $\omega_{\rm p}$  is approximately in the tens to hundreds of gigahertz range which matches with the time interval in the preceding paragraph \citep{ZhangF18}. On the other hand, the growth rate of the instability is $\propto t_{\rm q}^{-2}$  assisted by the self-excitation for sufficiently small time cycles \citep{gal+81}. Thus, a very short caviton passage time can be enough to produce excessive potentials observed in the ns shots of SGPs and probably existing in FRBs. To our knowledge, this is the first time the stimulated emission is associated with the Buneman instability and the relevance of these mechanisms for FRBs is discussed. The main  novelty of this research, the breather-type solitons and non-linear two-level system (see \S~\ref{ssec:nonlintwo}) are explained in the context of the stimulated emission to provide coherent radiation.

\begin{table}
	\begin{center}
		\begin{tabular}{|l|l|} \hline
			Parameter  & Value \\ \hline
			Bulk Electron density  & $n_{\rm e} = 10^{11} -  10^{14} \, {\rm cm}^{-3}$\\ \hline
			Bunched Electron density & $n_{\rm b} = 10^{7} -  10^{8} \, {\rm cm}^{-3}$ \\ \hline
			Temperature & $8 \times 10^6\, {\rm K}$ \\ \hline
			Magnetic field & $B_{\rm L} \sim 2-3\times 10^5$~G \\ \hline
			Cyclotron frequency & $\omega_{\rm B} < 840 $ GHz  \\ \hline
			Plasma  frequency &  $ 1\, {\rm GHz}< \omega_{\rm p} < 90\, {\rm GHz}$ \\ \hline
		\end{tabular}
	\end{center}
	\caption{Reference parameters of the plasma close to the current sheet.}  \label{tb:param}
\end{table}

In the meantime, the sensitivity to changes in wave frequency of the non-linear dispersion and the envelope soliton for electric field and polarization is very low compared with others; thus only the matching condition is required. Hence, the broadband frequency spectrum can be covered as long as they are matched by the self-excitation mechanism. The local oscillations of the breather-type soliton are usually characterized by short time duration, as short as a few cycles, residing in the traveling wave \citep{Herve19}. 

The observed narrower pulse width of the emission corresponds to smaller beam opening angles \citep{ZhangF18}.
By considering the relatively short time window ($\sim 1\,{\rm ns}$) of SGPs, the opening angle of the narrow  cone for coherent emission or angle of aperture,  $\theta$, could have a maximum value of $\theta_{\max}=2^{\circ}$, calculated as the one-tenth of the average pulsar inclination angle \citep{ZhangF18} where $\theta \ll \theta_{\max}/10^6$ radian due to very short-duration breather-type solitons complying with the numerical results \citep{Asseo+80,ZhangF18}. In the relativistic environment, a highly focused beam can have a much narrower tubular form with slanted emission direction \citep{dyks17} that emphasises the ``rare'' probability to match the line of sight. Relativistic particle velocities and higher frequency waves produce much smaller beam-width similar to narrower pulse effect \citep{mel-yuen16,lori-kramer12,ZhangF18}. Consequently, the velocity modulation or self-modulation at the boundary of the light cylinder close to current sheets, forms electron bunches that pass through a cavity-like resonator. These relativistic bunches are then accelerated/amplified by klystron-like structures with the evolution of ``rare'' matched conditions, e.g.\ ``self-excitation'' of the natural resonance modes. Thus, robust microwave emission in wild ambient plasma can pass through the outer space as strongly focused in a very narrow conic region.

If our model is correct we can predict novel multi-beam geometry inspired by the helix beam suggested by \citet{dyks17}. To represent the breather-type solitons of the helical vibrations/oscillations, we can propose a multi-beam model such that they originate from a very narrow tubular source. Furthermore, this model will help to match the discontinuities/anomalies in the observations such as drifting, nulling or double notch (profile moding) due to the single pulse assumption \citep{dyks17,basu+19}.

The extreme plasma lensing reported by \citet{mai+18} for the galactic millisecond pulsar B1957+20 indicates that radio pulses can be strongly
amplified by lensing in ambient plasma \citep[see also][for the case of B1744-24A]{bil+19}. The flux enhancement factors of up to 70--80 at specific frequencies by plasma lensing implies a possible similarity with the FRB phenomena \citep{mai+18}. If the $16.35\pm 0.18\,{\rm d}$  periodicity observed from FRB 180916.J0158+65 \citep{ami+20} is the orbital period of a binary system with an active pulsar, then this FRB can be considered as an extreme case of strongly amplified pulses from the above mentioned galactic millisecond pulsars where giant pulses are lensed through the ambient plasma provided by the stellar wind during periastron passage. The mechanism we propose here does not 
exclude the plasma lensing, but exists whenever a strong pulse travels through the medium.

\appendix 

\section{Coupling between the linear approximation and the non-linear population dynamics}
\label{sec:append}

Here we want to show how the approximation of the Gaussian probability distribution for the trapped fraction in \autoref{eq:trap}
\begin{equation*}  
\frac{n_{\rm tr}}{n_{\rm e}} = \frac{1}{\sqrt{2\,{\rm \pi}}} \int_{v_{\rm p} -v_{\rm tr}}^{v_{\rm p} + v_{\rm tr}}   {\rm e}^{-v^2/v_{\rm e}^2} \, {\rm d}v  \, ,
\end{equation*}
is equavalent to the differences of tangent hyperbolic functions 
\begin{equation}
\frac{n_{\rm tr}}{n_{\rm e}} \approx K_1 \left(\tanh(v_{\rm p} + v_{\rm tr}) - \tanh(v_{\rm p} - v_{\rm tr})  \right)
\end{equation}
where $K_1$ is a constant, as used in literature.
In \autoref{eq:popdiff} the magnitude of the polarisation vector $P$  can be approximated by its non-linear part (main envelope soliton) 
\begin{equation}
P \approx K_2 {\rm sech} (x)
\end{equation} 
where $x \equiv  (d_0 E_{\rm m}/2\hbar) \left( t- z/v_{\rm g}\right) $ is the argument (field-dependent) for polarization and $K_2$ is a constant. When this approximation is substituted in \autoref{eq:popdiff}, we obtain 
\begin{equation}
W \approx K_2 \tanh^2 (x).
\end{equation}
We thus show the same envelope shape is valid for both the linear and non-linear part as the tangent hyperbolic function of field-dependent variables. This is sufficient for the proof of the approximation we used in population inversion dynamics in \autoref{eq:trap} and \autoref{eq:popdiff}.

\section{Derivation of the nonlinear dispersion relation}
\label{sec:app-B}

In this section we sketch the derivation of the nonlinear dispersion relation given in \autoref{eq:n0} from \autoref{eq:main1} and \autoref{eq:main2} as explained in \S 5.6 of \citet{kraso08}.
One seeks solutions of \autoref{eq:main1} and \autoref{eq:main2} of the form
\begin{align}
E = N d_0 \mathrm{Re} {\cal E}(\xi) \exp( \mathrm{i} \Phi),  \\
P = N d_0 \mathrm{Re} {\cal A}(\xi) \exp( \mathrm{i} \Phi)
\end{align}
where 
\begin{equation}
\Phi \equiv \omega t - k z, \qquad \xi \equiv \omega_{\rm p} (z - u t)
\end{equation}
\citep{kra65}. These equation are nonlinear wave envelope equations. Here $u$ is the wave velocity. 
After substituting the equations above into \autoref{eq:main1} and \autoref{eq:main2}, and using the inequalities given in \autoref{eq:ineq} which allows one to ignore the second derivatives of the amplitudes we obtain the ordinary differential equations
\begin{align}
2{\rm i} (n- \beta){\cal E}' + (n^2 - 1 ){\cal E} &= 4{\rm \pi} {\cal A}, \\
-2{\rm i} \beta {\cal A}' + \delta {\cal A} &= \frac{q^2}{4{\rm \pi}} \sqrt{1 - {\cal A}^2}\, {\cal E}
\end{align}
 where $\beta = u/c$. If one eliminates ${\cal A}$ from the differential equations one obtains the nonlinear equation for the amplitude of the dimensionless complex field 
 \begin{equation}
 4\beta(n-\beta) {\cal E}'' + 2 {\rm i} \Psi {\cal E}' + \delta (n^2 - 1){\cal E} 
 = \left( \frac{q}{4{\rm \pi}} \right)^2 \Lambda^{1/2} {\cal E}.
 \label{eq:ara}
 \end{equation}
 where
 \begin{equation}
 \Psi =  \delta (n- \beta) - \beta (n^2 - 1 )
 \end{equation}
 and 
 \begin{align}
 \Lambda = &16 {\rm \pi}^2 - 4 (n- \beta)^2 |{\cal E}'|^2 - (n^2 - 1)^2|{\cal E}|^2 \\ \nonumber
    & - 2 {\rm i} (n-\beta)({\cal E}^*{\cal E}'-{\cal E}^*{\cal E}^{*\prime})  
 \end{align}
where `$*$' denotes the complex conjugate.
To simplify further we can assume $\Psi=0$ which implies
\begin{equation}
\beta = \frac{n}{1 + (n^2-1)/\delta}
\end{equation}
and ${\cal E}$ is real which corresponds to a wave with resonant group velocity.
We now define
\begin{equation}
y \equiv \frac{\cal E}{4{\rm \pi}} |n^2 - 1|^2, \quad Q \equiv \frac{q^2}{\delta(n^2-1)}, \quad \tau \equiv \frac{|\delta|}{2\beta}\xi 
\end{equation}
where $\tau$ is the new time variable. These allow one to recast \autoref{eq:ara} as
\begin{equation}
y'' + y = Q \sqrt{1 - y^2 - y'^2}\, y
\end{equation}
where the primes denote derivatives with respect to $\tau$.
A soliton solution satisfying $Q>1$ (recall $Q \simeq \delta/(n^2-1)$ is satisfied easily) is
\begin{equation}
y = \frac{2}{Q} \sqrt{Q-1} \, {\rm sech} (\sqrt{Q-1}\, \tau)
\end{equation}
which is equivalent to \autoref{eq:n0} and results with \autoref{eq:group}.

\section*{Acknowledgements}
KYE acknowledges support from T{\"U}B{\. I}TAK with grant number 118F028.

\footnotesize{
\bibliographystyle{mn2e}
\bibliography{frb,crust,giant,radiation,sheet,plasma} 
}

\end{document}